\documentclass[%
 reprint,
 amsmath,amssymb,
 aps,
 prl,
]{revtex4-2}

\usepackage{graphicx}
\usepackage{dcolumn}
\usepackage{bm}
\usepackage[utf8]{inputenc} 
\usepackage[T1]{fontenc} 
\usepackage{amsmath}
\usepackage{amssymb}

\begin{document}


\title{Spherical and Deformed Shell Effect Competition in Quasifission of Superheavy Nuclei}

\author{Richard Gumbel}
\email{gumbelri@msu.edu}
 \altaffiliation[Also at ]{Physics Department, Michigan State University.}
\author{Kyle Godbey}
\email{godbey@frib.msu.edu}
\affiliation{Facility for Rare Isotope Beams, Michigan State University, East Lansing, Michigan 48824, USA}

\date{\today}

\begin{abstract}
Quasifission, along with fusion-fission, represent the two most likely reaction outcomes to occur post-capture in collisions leading to superheavy nuclei.
As such, understanding these mechanisms and how they relate to one another is
key to understanding the intricate dynamics that drive the formation (or dissociation) of the nascent compound nuclei formed in fusion reactions.
This understanding directly translates to a more informed picture of suitable reaction partners and can provide vital information for experimental efforts to study the physics and chemistry of superheavy elements.
In this work we report results from time-dependent simulations of $^{48}$Ca + $^{238}$U and $^{50}$Ti + $^{236}$Th reactions at incident energies just above the Coulomb barrier with a focus on the quasifission process that prevent the formation of a fully equilibrated $^{286}$Cn compound nucleus.
We study these reactions systematically and consider a wide range of initial configurations to extract a robust estimate of primary fragment yields for the quasifission process.
Multiple preferred exit channels are observed, with both spherical and deformed shell effects in the heavy and light fragments driving contributions to the production yields depending on the initial configuration of the system.
Orientation effects of the deformed actinide targets are found to be a primary driver of which exit channels are populated. 
Furthermore, the impact of moving away from a doubly-magic projectile is explored with implications towards the reactions considered for current and future superheavy searches.

\end{abstract}

\maketitle



{\it Introduction ---} Efforts to extend the periodic table through the synthesis of new transactinides are underway in experimental facilities around the world~\cite{smits2024}.  The successful creation and study of these superheavy elements (SHE) stands to impact a wide range of research fields.  In quantum chemistry they present an opportunity to study strong relativistic effects on the electrons in the atom~\cite{schadel2014}. In astrophysics, the production of transient SHE by nucleosynthesis remains an open question -- whether or not these elements will be produced in any appreciable quantity in stellar environments as a result of the rapid neutron capture (r-process)  in the extreme conditions of super novas and neutron star mergers and the impact they will have on abundances is currently not known~\cite{aprahamian2024, GORIELY2015}.   From a nuclear physics perspective, the study of superheavy nuclei (SHN) offers deep insights into our basic understanding of nuclear structure and dynamics. The most fundamental of which is the question of stability and existence in the far reaches of the nuclear landscape, where predicted shell closures in neutron rich SHN point to the existence of a yet-to-be-confirmed region of increased nuclear stability \cite{Oganessian_2012, oganessian2015, smits2024}.  Theoretical predictions vary on the possible increase in the lifetimes exhibited by isotopes in this region, though even a modest increase in lifetime can have profound impacts on nucleosynthesis or our ability to produce meaningful amounts for further study.

SHN are identified by the decay properties of a formed evaporation residue -- evidence that an equilibrated compound nucleus has formed in an excited state and then shed some of its energy via particle emission. 
These SHN that are left are still unstable, though with a half-life much longer than the typical nuclear dynamics timescale of $~10^{-21}$~s. 
Quasifission reactions present a substantial hindrance to SHN formation \cite{Sahm1984}.  In such a reaction, capture occurs between two colliding nuclei with contact times on the order of $10^{-21} - 10^{-20}$ seconds \cite{knyazheva2014, ray2015}: sufficient for substantial transfer of nucleons between collision partners as well as full dissipation of kinetic energy~\cite{oberacker2014}.
Before a fully equilibrated compound nucleus can be formed, however, the nascent pseudo-compound will scission~\cite{hinde2016}. This is primarily due to the fact that a transfer of charge takes place much sooner than bulk mass transfer can occur. For the entrance channel of medium-to-heavy mass systems, charge products $Z_1 Z_2 \geq 1600$~\cite{kayaalp2024} lead to an extremely strong Coulomb repulsion that erodes the conditions favorable to fusion of two nuclei.  As a consequence, the evaporation residue cross sections associated with SHN formation, are suppressed by quasifission (along with fusion-fission) to the order of pico-barns~\cite{vardaci2019}. Such  competition between reaction outcomes demonstrates the need for a thorough understanding of quasifission reactions, and is particularly relevant for experimental efforts to extend the periodic table further past its current limit of the element oganesson ($Z=118$).

The collision and capture of two nuclei leading to quasifission exhibits some of the same characteristics to those of fusion-fission reactions, and the two processes can be difficult to disentangle experimentally~\cite{godbey2020}.  While quasifission is typically a much faster process, the upper limit of the time scale of quasifission can approach the average minimum associated with fusion-fission.  In addition, both reactions are impacted by the presence of quantum shell effects in the prefragments and compound system~\cite{godbey2019,simenel2021a} which impact the observed final distributions of nuclear mass in a similar way.  For fissioning nuclei, shell effects tend to drive the mass fragmentation away from symmetry.  In quasifission, shell effects function effectively as a brake to the mass equilibration process \cite{simenel2021a}.  The shell effects present in quasifission then hamper the intermediate stage by which the formation of a fully equilibrated compound nucleus would take place. 
The quasifission exit channel is driven, largely, by the pseudocompound nucleus' angular momentum and excitation energy which leads to a memory of the entrance channel for quasifission, which is absent from fusion-fission.

In this paper, we present the results of simulations of $^{238}$U with $^{48}$Ca, as well as $^{50}$Ti and $^{236}$Th, both of which combine to form the superheavy nucleus $^{286}$Cn. 
These are examples of 'hot fusion' reactions, where a neutron rich projectile collides with an actinide target, which experimentally have met with success in synthesizing elements $Z=114-118$ (\cite{roberto2015, gates2024}).
Each system is studied through the use of time-dependent Hartree-Fock (TDHF) methods for a range of initial angular momenta and orientations of the deformed $^{238}$U and $^{236}$Th targets.
These systems are studied at incident energies of 5\% above the Coulomb potential barrier.
Exploring these initial configurations systematically gives a detailed picture of the most likely reaction outcome for each initial state.
At a given energy, the reactions vary from deep inelastic collision for reactions with large impact parameters to the capture of target and projectile for small.
This process also displays a strong dependence on the relative orientation of the actinide target during central collisions. 

Studying these collisions then is crucial to characterizing how conditions of equilibrium develop in systems of complex nuclei \cite{viola1987,cook2023colliding}.  In addition to better understanding entrance channel effects on the SHE formation process, we also aim to explore quasifission’s potential as a unique probe of out-of-equilibrium nuclear systems. That is, what information quasifission can provide on mass/isospin equilibration, timescales of interactions, and how both are impacted by deformations of nuclei and the dynamical evolution of shell effects.


{\it Methods ---} Density functional theory is our tool of choice for examining the quantum many-body dynamics of quasifissioning nuclei.  Specifically, the self-consistent mean field theory of the Hartree-Fock equations paired with Skyrme's interaction.  (For detailed derivations of TDHF, some relevant resources include \cite{vautherin1972, sekizawa2015, suckling2011, simenel2018-zy, nakatsukasa2016}.)  Reactions were modeled using the VU-TDHF3D code developed at Vanderbilt University by A.S. Umar et al. The program implements the static and time-dependent Hartree-Fock frameworks for interactions of Skyrme-type density functionals, in three dimensional Cartesian coordinates and with no assumptions of any spatial symmetries.
This flexibility makes it well suited for modeling both structure and reactions, though the standard caveat is that the result within the TDHF framework should be interpreted as the most probable outcome of a nuclear reaction.
TDHF will fail to capture the full distribution of observables that can occur from a given set of initial conditions.  With this limitation in mind, density functional theory remains a powerful predictive tool, particularly in its ability to explore the dynamics of systems for a wide range of nuclei.   

\begin{figure}
\includegraphics[width=0.49\textwidth]{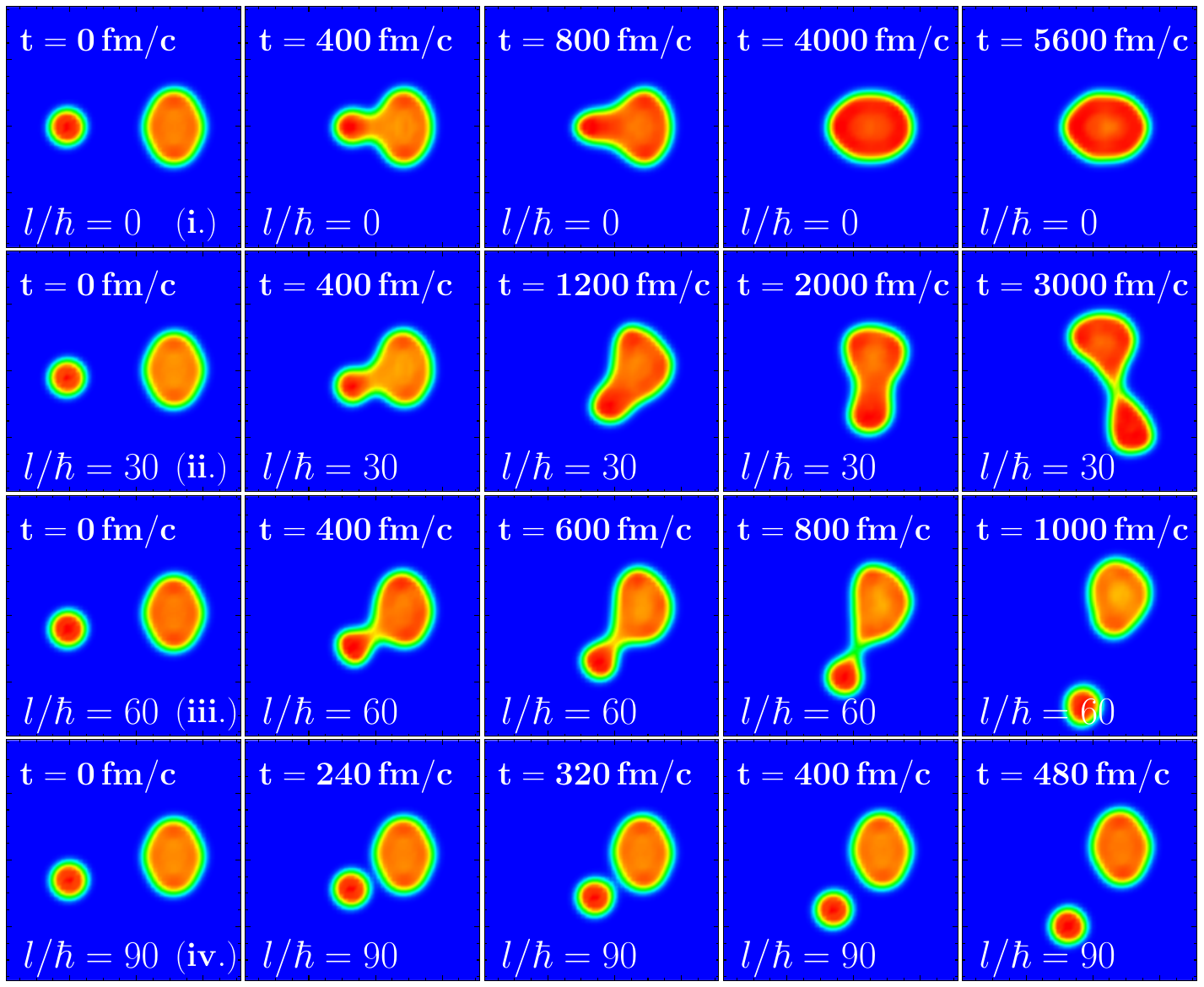}
\caption{Density plot of $^{48}$Ca + $^{238}$U at four values of orbital angular momenta: (i.) Central collision, at $l = 0 \,\hbar$ resulting in capture and no observed scission, (ii.) Capture and immediate quasifission occurs $l = 30 \, \hbar$, (iii.) Few nucleon transfer/short timescale of quasi-elastic collision at $l = 60 \, \hbar$, (iv.) No transfer and no contact at $l = 90 \, \hbar$.  The uranium target is rotated $\beta = 90^\circ$ along the collision axis, and the center of mass energy is fixed at $E_{CM} = 203.53$ MeV. TDHF calculations performed with the Sly4d parameter set.}
\label{fig:Ca48U238_panelplot}
\end{figure}

\begin{figure*}[htb!]
    \includegraphics[width=0.9\textwidth]{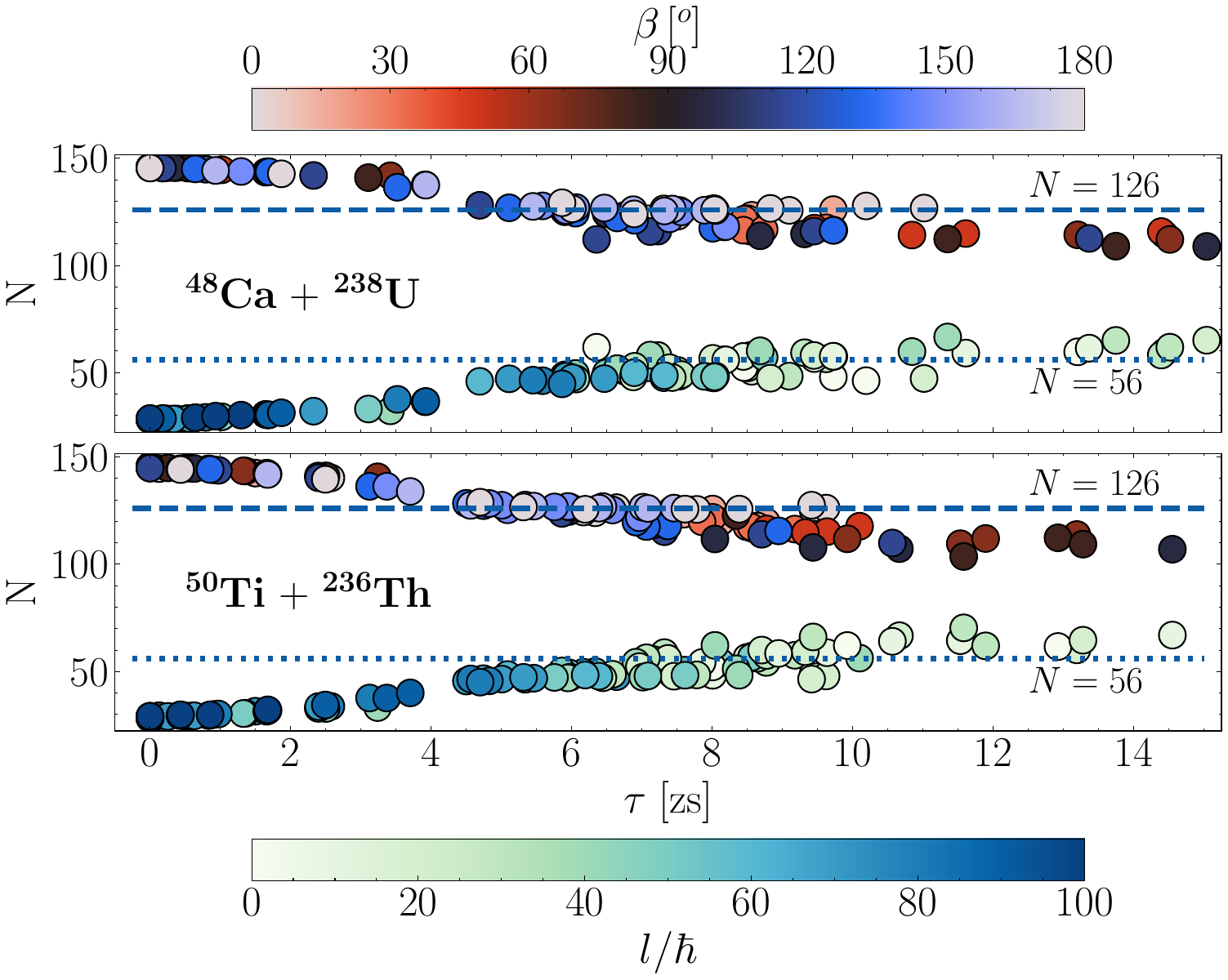}
    \caption{Fragment neutron numbers vs. contact time($\tau$) for both systems: $^{48}$Ca + $^{238}$U ($E_{CM} = 203.53$ MeV), $^{50}$Ti + $^{236}$Th ($E_{CM} = 219.27$ MeV). Points corresponding to heavy fragments (N $> 87$) are color coded to indicate the initial orientation of the actinide target in the entrance channel; the lighter fragment points (N $< 87$) shows the initial orbital angular momentum of the collision as a function of impact parameter $l = \sqrt{2mE_{CM}}b$. Calculated with Skyrme interaction parameter set SLy4d.}
    \label{fig:N_v_ctime}
\end{figure*}

We first performed individual static Hartree-Fock calculations for the nuclei of interest, $^{48}$Ca, $^{50}$Ti, $^{238}$U, and $^{236}$Th, on a 3D Cartesian grid of dimensions $56 \times 56 \times 36$ fm.  After generating the ground state solutions
we then carried out a series of time-dependent Hartree-Fock runs. 
For systems with pairing correlations, the frozen occupation approximation was used for the dynamical evolution.
For each collision, within the center of mass frame, the relative orientation of the actinide target to projectile could assume a range of values from 0$^\circ$-165$^\circ$ along its $\beta$ Euler angle, spaced at fifteen degree intervals.  At each of these fixed initial orientations $\beta$, 11 different values for angular momentum were examined, $l=0-100 \, \hbar$ at intervals of $10 \, \hbar$.  The initial center of mass separation between collision pairs was 25 fm.  For incident energies, we chose values of approximately five percent above each system's projected Coulomb barrier, calculated using a Bass potential model. For $^{48}$Ca + $^{238}$U, the center of mass energy used was $E_{CM} = 203.53$ MeV; while for $^{50}$Ti + $^{236}$Th, $E_{CM}= 219.7$ MeV. This resulted in a total of 132 TDHF trajectories for each target-projectile combination.
Ideally, a finer $l$~mesh and full angular integration of the deformed target nucleus would be performed, but the values chosen give good coverage of the reaction systematics. 


{\it Results ---} We begin with an overview of the types of reactions observed in the TDHF calculations of this study and their associated time scales.  Fig~\ref{fig:Ca48U238_panelplot} shows nuclear densities evolving in time for the system $^{48}$Ca + $^{238}$U where we've rotated the prolate deformed uranium target ($\beta = 90^{o}$), and examined collisions for four different values of angular momentum.  At $l = 0 \, \hbar$, head-on, capture occurs, and the system fuses, i.e. its relative energy of motion is converted to internal excitations and its collective degrees of freedom without scissioning. The introduction of angular momentum with $l = 30 \, \hbar$ in the second row of figures, shows capture between the nuclei, followed swiftly by quasifission, fragmenting into $^{98}$Y ($Z=39$) and $^{187}$Ta ($Z=73$).  The sequence in the third row of figures ($l = 60 \, \hbar$) takes place on an even quicker timescale, producing $^{51}$Sc ($Z=21$) and $^{235}$Pa ($Z=91$).  The few nucleon transfer and brief contact time indicates a deep-inelastic collision, having surpassed the threshold of critical angular momentum beyond which quasifission ceases to occur for a given relative orientation (in this case $l_c \sim 50 \, \hbar$).
Finally, the fourth row of figures ($l = 90 \, \hbar$), shows a quasielastic collision with no transfer of nucleons and minimal energy dissipation.

With this broad categorization of the different reactions that may result within a given system in mind, we now examine the partition of nucleons as a function of contact time in Fig.~\ref{fig:N_v_ctime}.
Additionally, we have further color coded each trajectory with the system's initial angular momentum and initial relative orientation.  Figure~\ref{fig:N_v_ctime} shows the final distribution of neutrons vs duration of contact between collision pairs $^{48}$Ca$+^{238}$U and $^{50}$Ti$+^{236}$Th (upper and lower panels respectively). Contact time is defined as the interval of time that the overlap in the total density is equivalent to roughly half nuclear saturation density $\rho = 0.08 \, \text{nucleons}/\text{fm}^3$.  

For central collisions ($l =0 \, \hbar$), the orientation of the prolate deformed actinide target had a distinct impact on the duration of contact.  Either spherical projectile colliding with the tip of its prolate deformed actinide target resulted in immediate quasifission, with contact times $\tau \leq 10 \, \text{zs}$.  A  rotation $\beta=90^\circ$ of the uranium or thorium nuclei along the collision axis, resulting in projectile contact with the side of the target, yielded substantially longer contact times (as seen in the first row of images in fig \ref{fig:Ca48U238_panelplot}) and thus a larger window for equilibration to take place.  In the case of either projectile, a 'side-to-side' collision \textit{did not} result in scission within the timeframe of our modeled collisions (upper limit $\tau \approx 6000$ fm/c or $20$ zs). This orientation dependence for contact times is consistent with experimental results (\cite{hinde1995}), and previous TDHF studies of heavy ion collisions involving $^{238}$U (\cite{umar2016a, kohley2015}). \\

\begin{figure}
\includegraphics[width=0.48\textwidth]{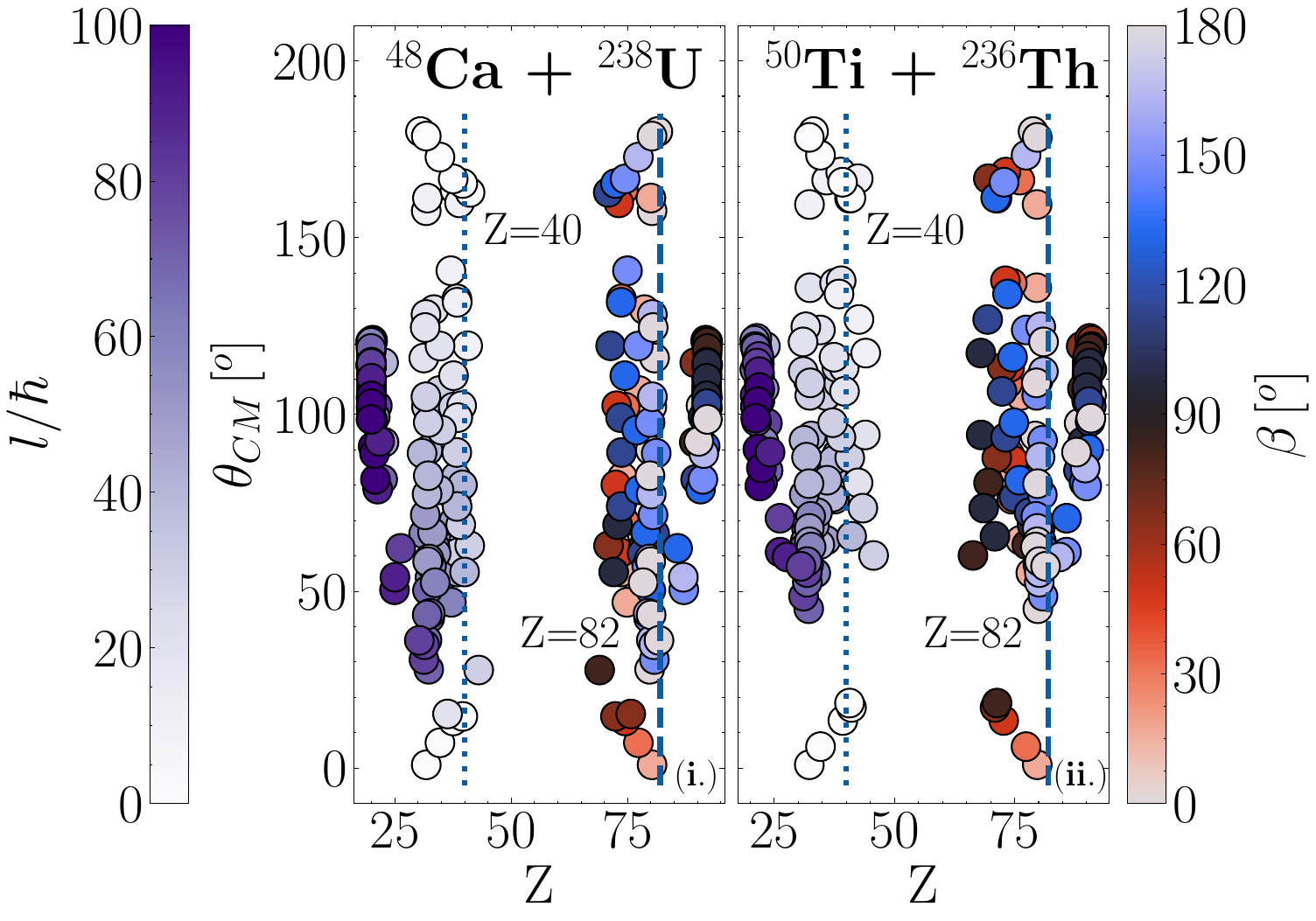}
\caption{$\theta_{CM} \, \text{vs} \, Z$ for both paths to $^{286}$Cn: $^{48}$Ca + $^{238}$U, $^{50}$Ti + $^{236}$Th. We see quasi-elastic collisions for large $l$. Calculated with SLy4d.}
\label{fig:theta_v_proton}
\end{figure} 

For intermediate values of ($10 \, \hbar \leq l \leq 70 \, \hbar$), paired with a range of orientations $\beta$, the impact of shell effects becomes readily apparent.
We see that collisions with values of mid-range values of $l$ take on one of two identities: a quicker exit corresponding to heavy fragments with $N\approx 126$ for tip-like collisions ($\beta < 30^\circ$ and $\beta > 150^\circ$) and a slower process more driven by deformed shell effects in the light fragment for side-like collisions ($30^\circ \leq \beta \leq 150^\circ$).
The difference in contact times is not wholly unexpected given the more compact shapes formed by the side-like collisions, though even the few tip-like counts that have $\tau > 6$ zs remain at $N\approx 126$ and do not equilibrate further.
Considering now the proton number of the fragments, Fig.~\ref{fig:theta_v_proton} shows the separation angle $\theta_{CM}$ as function of final charge fragments for our two systems, $^{48}$Ca + $^{238}$U, $^{50}$Ti + $^{236}$Th.
Both of these, showed little, if any, formation of lead isotopes, peaking mostly around mercury ($Z=80)$ and germanium ($Z=32$) for the tip-like collisions.
As the initial orientation approaches the side configuration, there is a stronger tendency toward symmetry with the most equilibrated systems producing isotopes around zirconium ($Z=40$) as the light fragment.
In the case of the $^{48}$Ca induced reaction, this manifests as a, mostly, two peak structure that corresponds to the two channels that emerge for neutrons in Fig.~\ref{fig:N_v_ctime}.
For $^{50}$Ti + $^{236}$Th however the charge production yields are much more dispersed and, for the side-like collisions, appear to be primarily driven by the neutron number of the light fragments.
This subtle difference in quasifission fragment production is likely a result of the doubly magic nature of the $^{48}$Ca projectile, though further reaction studies with $^{50}$Ti and nearby systems are certainly warranted to verify this behavior.
Such heavy fragment charge yields peaking shy of $Z=82$ were also found in a quasifission study of $^{48}$Ca + $^{249}$Bk~\cite{godbey2020}, though in that case only a single peak was observed and the significant spreading seen in the charge yields in Fig.~\ref{fig:theta_v_proton} was not present.
The study in Ref.~\cite{umar2016b} includes both $^{48}$Ca and $^{50}$Ti on $^{249}$Bk, though the limited systematics make it difficult to see if the same conclusions can be drawn for the heavier $^{249}$Bk target.

\begin{figure} [htb]
\includegraphics[width=0.5\textwidth]{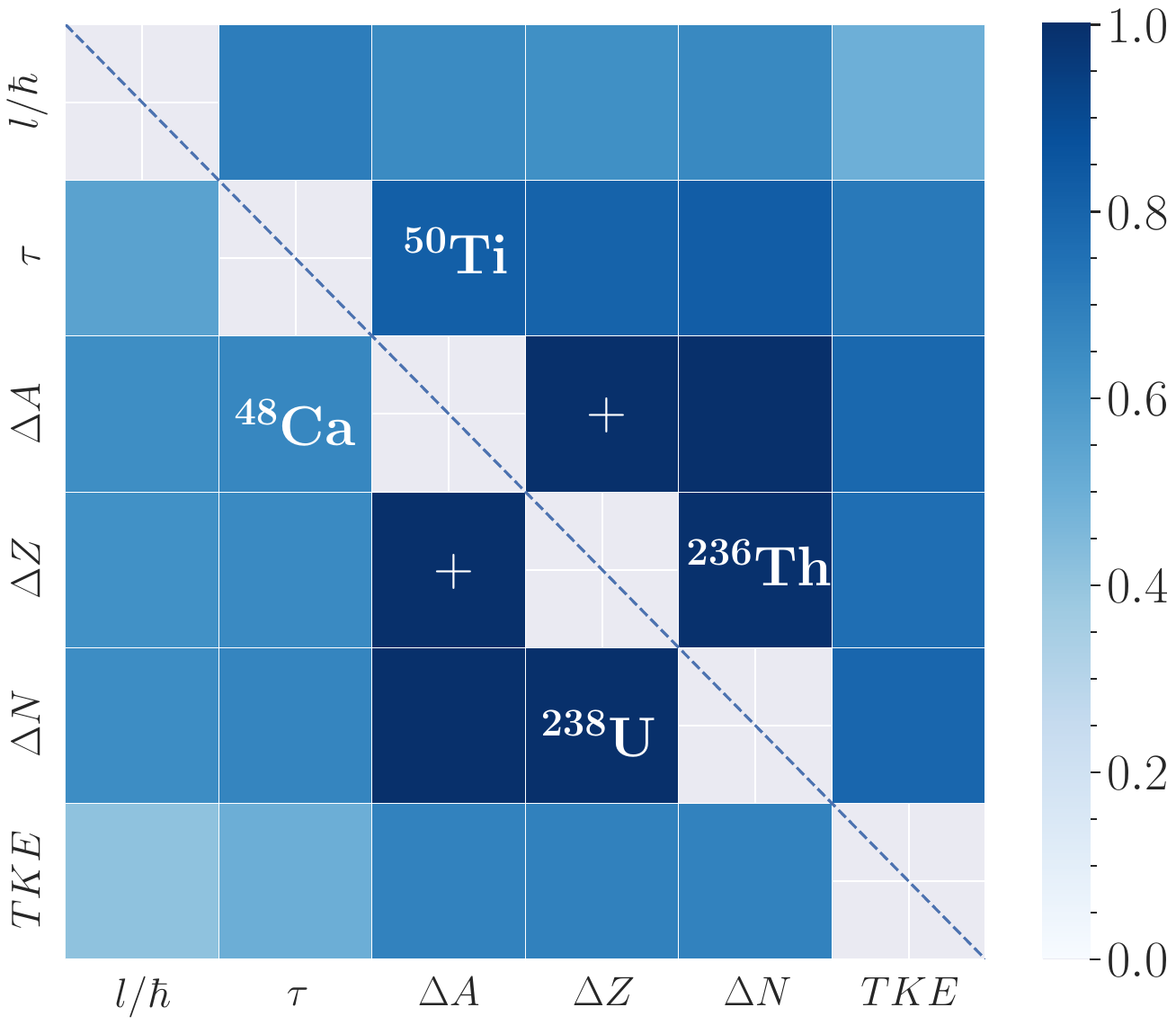}
\caption{Coefficients of determination for TDHF collisions of $^{48}$Ca + $^{238}$U at $E_{CM} = 203.53$ MeV, 
(lower left) and $^{50}$Ti + $^{236}$Th at $E_{CM} = 219.27$ MeV (upper right). Categories are contact time $\tau$, angular momentum $l/\hbar$, final kinetic energy of the system $TKE$, and final equilibration of nucleons per collision $\Delta X = (X_{\text{target}}(\tau_{f}) - X_{\text{projectile}}(\tau_{f})) / (X_{\text{target}}(\tau_{0}) - X_{\text{projectile}}(\tau_{0}) )$.}
\label{fig:COD}
\end{figure}

While the two systems display significant similarities in reaction outcomes, e.g. a strong presence of shell effects in the transfer of nucleons during quasi-fission, there are subtle differences in the dynamics of that mass equilibration.  While the competition between spherical and deformed shell effects occurs in both systems, contact times where we see evidence of these shell closures persist the longest for $^{48}$Ca on $^{238}$U.  The minimum time for quasifission events for both systems is roughly the same time ($\tau \approx 4.0 $ zs).
Similarly, the behavior as a function of $\beta$ and $l$ are roughly the same as well.
For the $^{50}$Ti on $^{236}$Th system however, the `shell stabilized' equilibration ends around $\tau \approx 10$ zs, after which we see a renewed trend of mass equilibration from the lighter fragment to the heavier.
This time spent at an equilibration stopping point is interesting in that, although both systems have the same configuration of neutrons in the entrance channel ($N_{\text{projectile}} = 28$, $N_{\text{target}} = 146$), once past the initial decay path of 'fast quasi-fission' our titanium system approaches isospin equilibration sooner than calcium on uranium.
Fig~\ref{fig:COD} depicts coefficients of determination (COD) for the reaction observables for the $^{48}$Ca + $^{238}$U (lower left) and $^{50}$Ti + $^{236}$Th (upper right) systems. Here, we examine the square of linear correlations in reaction observables, such as contact time $\tau$, nucleon and mass equilibration $\Delta X$, final kinetic energy $TKE$--and entrance channel considerations such as angular momentum, $l/\hbar$.  We see mutual linear dependence in the variance of these quantities for both systems.  As we'd expect there is a strong correlation between angular momentum and contact time, and thus degree of mass equilibration.  Consistent with the results of figures \ref{fig:N_v_ctime} and \ref{fig:theta_v_proton}, correlations are most pronounced for the system $^{50}$Ti + $^{236}$Th, which overall displays a larger sensitivity to variance in equilibration measures being accounted for by the contact time.
This is a result of the broader spread of fragment yields for the $^{50}$Ti system, as tightly peaked fragment production in Fig.~\ref{fig:N_v_ctime} would manifest as a straight line and, thus, lower correlation.

{\it Summary ---} In an era of superheavy research where evaporation residue cross sections are vanishingly small and superheavy searches can be expected to take months to years of beamtime, there is clearly a strong motivation to better understand the intricacies of the heavy-ion reaction dynamics and the subsequent impact of different projectiles on fusion probabilities at varying beam energies.
Laboratories around the world are actively exploring alternatives to $^{48}$Ca induced reactions for this reason, though only recently have groups shown success with $^{50}$Ti~\cite{gates2024} and $^{54}$Cr~\cite{Oganessian2024-eu} beams for the production of superheavy nuclei.
These efforts are vital for providing more data for understanding the evolution of fusion probabilities as one moves across the nuclear chart and, more broadly, provides key benchmarks for the structure and dynamics of atomic nuclei.
Due to the inherent complexity in describing near-barrier reactions of heavy nuclei, fusion and related processes remain an excellent probe of models that attempt to self-consistently describe the quantum many-body problem~\cite{cook2023colliding}.
We have explored these questions of entrance channel effects through a large-scale systematic study of two reactions leading to the same superheavy nucleus, $^{286}$Cn. The reactions were chosen explicitly to also explore the impact of the choice of projectile, given the recent successes at LBNL with $^{50}$Ti beams. The TDHF method employed has been well used within the community for heavy-ion reactions at near barrier energies, and is well suited for the reaction observables considered here.

The most striking result is in the apparent competition between spherical and deformed shell effects driving quasifission fragment production for both systems.
Depending on the interplay between the initial orientation of the actinide targets and the centrality of the collision, the compound system tended towards two primary outcomes --- one driven by spherical shell gaps in the heavy fragment near $N=126$ and another driven by deformed shell gaps near $N=56$.
The orientation effect in particular seems to be most tightly correlated with the reaction outcome, suggesting that the side-like collisions leading to $N=56$ events will be more likely in an experimental context.
We also observed that the charge yields of the $^{50}$Ti induced reactions were significantly broader than that of the $^{48}$Ca induced reactions, though the primary driver of fragment production for both systems was still the initial target orientation.

While the primary motivation of the current work is focused on quasifission as a reaction channel that hinders fusion, one can also imagine exploiting these shell effects to produce novel superheavy nuclei or, more realistically, neutron-rich nuclei far from stability.
By exploiting these preferred fragment particle numbers, one may be able to control fragment production out of equilibrium in a process known as \textit{inverse} quasifission.  That is, quasifission reactions where the transfer of nucleons flows from the lighter collision partner to the heavier \cite{kozulin2014b, zagrebaev2005, sekizawa2015}.
Experimental confirmation of this process and the resulting data would act as an excellent constraint for nuclear models and potentially open the door for studies of exotic nuclei that are currently difficult to access at leading rare isotope beam facilities.

{\it Acknowledgments ---} We would like to thank Witold Nazarewicz for providing feedback and discussions on the present work. This work was supported by the Department of Energy under Award Numbers DOE-DENA0004074 (NNSA, the Stewardship Science Academic Alliances program) and DE-SC0023175 (Office of Science, NUCLEI SciDAC-5 collaboration).

\newpage

\bibliography{references}

\end{document}